# Mobile Learning Culture and Effects in Higher Education

Seibu Mary Jacob and Biju Issac, *Member, IEEE*

*Abstract*—Mobile learning through wireless enabled laptops (say, within a university campus) can make use of the learning management system that is already available through internet or intranet. Without restrictions within the four walls of computer labs or library, students can now access the learning resources anywhere in the campus where wireless access points or hotspots are located. We briefly investigate on the mobile learning benefits and eventually an analysis of the student perceptions on mobile learning is presented through a survey, to validate the m-learning benefits.

*Index Terms*—Mobile learning, Wireless networks, Student survey.

## I. INTRODUCTION

The technological era has gone far in terms of internet technology and thus ushered in e-learning via computers. Online courses in universities as well as courses with online resources became the state-of-the-art technology in e-learning. The immobile computers in the computer labs on-campus offered a digital library at the student's finger tips. But then these online resources subscribed to by the university could be accessed only in campus. Here's where portable PCs came in to the rescue, and then the concept of wireless LANs. A wireless capability gives the institution a straight forward, cost effective solution to maximize all the benefits of the educational network. The availability of laptops which connect to the hotspots in campus removed the restriction from the campus lab PCs [1].

Higher education campuses can easily become fertile ground for wireless LANs fueled by the explosive adoption of mobile devices among students and faculty [2]. The recent affordability, power and usability of laptop computers have begun a trend towards portable computing in education to meet this need [3]-[4]. Keegan in his latest book provides systematically the benefits, problems and recommendations to enhance learning in mobile environments [5]. It is a world of information at the fingertips of the present generation. Brown talked of this as information navigation [6]. And then came in the constructivist approaches bringing in the concept of Communities of Practice (COPs). A Community of Practice perspective sees learning in more informal settings taking place as a by-product of joining a group of practitioners and having a legitimate, peripheral participation in one or more aspects of the practice being carried out by the experts [7].

## II. BENEFITS OF MOBILE LEARNING IN HIGHER EDUCATION

The benefits of m-learning (mobile learning) can be felt at the distinct levels as given below. They have some commonalities with benefits of e-learning approaches, as m-learning is a subset of e-learning. (1) *Easy Access* – knowledge is delivered on-demand, with updated information within the precincts of the m-learning campus, (2) *Options for Self-study* – the flexibility of m-learning enables participants to learn at their own time and pace even more compared to the fixed PC access. Hence the amount of information retained from the training is often greater, which results in increased information retention, (3) *Evaluation and Feedback* – *a*ssessment tools can be included into the m-learning or e-learning packages to monitor student's progress, and produce detailed usage reports. This can be given as feedbacks to students or learners, (4) *Access of Online Repository* – the online materials accessed through m-learning system offers a place for the lecturers and students to interact frequently. Learners have access to a stored repository of knowledge and information like the digital course materials and a host of other online digital libraries for assignments and exams. (5) *Communities of Practice* – the three elements of a COP are a domain, a community and a practice and the theory behind is that learning involves participation in a COP. Most COPs meet online and m-learning makes this click well [7]-[8].

## III. ANALYSIS OF STUDENT PERCEPTIONS ON M-LEARNING

A university survey was conducted to explore and analyze the factors crucial in overcoming the possible hindrances of m-learning implementation in higher education. Student perceptions of m-learning may be influenced by specific individual variables. The variables taken into consideration in this study were gender, course of study and attitudes to new technologies. Research has indicated that men are strongly influenced by perceptions of usefulness in technology usage decisions. But women are more attracted to the ease of use. Men and women focus on different aspects of using computers [9]. The authors felt it's worthy that the theories of technology acceptance be considered in studies of this sort. Rogers speaks of five different adopter categories in his description of a general framework of technology acceptance

Seibu Mary Jacob is a part-time PhD student in University Malaysia Sarawak, Malaysia. Biju Issac is with the School of IT and Multimedia, Swinburne University of Technology (Sarawak Campus), Malaysia who is also an IEEE ComSoc, EdSoc and CommSoc member. Contact email addresses: {sjacob, bissac}@swinburne.edu.my

Publisher Identification Number  1558-7908-IMCL2007-01

within the theory of diffusion of innovations. The five adopter categories–innovators, early adopters, early majority, late majority and laggards are regarded in this study [10]. SPSS software was used for analysis. The three specific objectives of the survey were: (1) to explore students' general attitudes to e-learning through wireless networking (or mobile learning) on campus; (2) to analyze the relationship between the attitudes in (1) and specific background factors like gender, course of study, attitudes to new technologies; (3) to explore the most important advantages and disadvantages that the students anticipate in the context of mobile learning. The scope of the study conducted in 2006, included students who have been exposed to wireless networks in the university environment during one semester. The questionnaire was distributed to a large sample of 250 students across Business and Engineering streams in a Malaysian university. The response rate was 76%, thus the analysis was based on the 190 responses received.

*A. The Methodology and Results of the Survey*

The first objective, namely the students' attitudes to m-learning was measured using seven closed questions. The statements hypothesized in the questions were:

1. Wireless networks offer seamless access to digital information, and hence is a boost to this information age.
2. The use of the wireless network can increase flexibility of access to resources (like Black Board website, slides, notes, library journal access etc.) in my studies.
3. The wireless networks are not generally very secure and so I wouldn't want to use it when I can use desktop PCs.
4. The use of the wireless network can improve communication with teachers and tutors.
5. The use of the wireless network can improve the learning (pedagogic) value of the courses and the courses are more recommendable to others.
6. With wireless network I do not need to depend on library PCs or lab PCs. Accessing of internet for working on assignments within University is a lot easier.
7. Do you prefer mobile phone to be used for mobile learning (since it can access web pages)?

It was noted that statement 3 was a negative statement unlike the others which were all positive statements. Statements 1-6 were given alternatives based on the Likert scaling on a scale of 1-5, where 5 represents 'I agree totally' and 1 represents 'I disagree totally'. As shown in the figure 1, the majority of the responses for the questions on the use of the wireless networks for m-learning have shown a mean close to 4 (I agree to a large extent). The best mean was 4.2 in response to the statement "wireless networks increases flexibility of access to resources in learning" and the worst mean 3.4 in response to the statement "wireless networks can improve communication with teachers and tutors". The implication was that the students agreed to a large extent to the easiness, flexibility, assistance, improved communication offered by the mobile learning platform. The statement 7 offered the alternatives: 'Yes' and 'No'. The responses to statement 7 showed the majority of students responded 'No (as the screen size is small)' to the usage of mobile phones for mobile learning against the alternative 'Yes (as I can at least access some information)'. Here it was noted that the Engineering students have been on the negative side than their Business counterparts. It could be indirectly inferred that the response is an indication of the technical know-how that the Engineering students have over Business students, since they preferred a laptop based network communication over mobile phone communication for full fledged web-based mobile learning that has heavy learning contents.

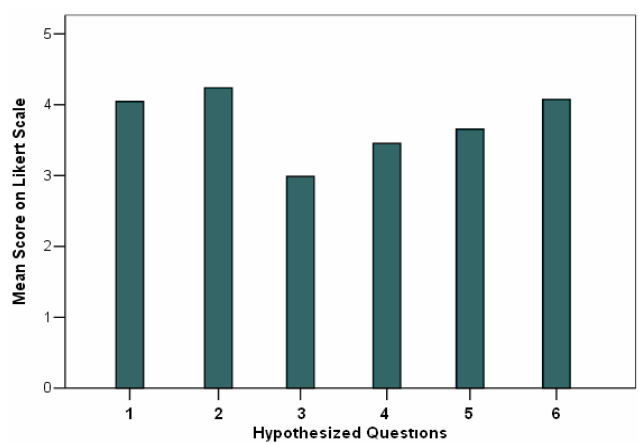

Fig 1. The mean Likert scale score for the student attitude towards mobile learning.

Under the second objective, a factor analysis was performed on the closed ended questions to show the inter-relationship between the questions. Then a multiple regression analysis was attempted using the variable 'Attitude'. This was an index formed by summing the responses to the statements 1-6 for each individual. The response 'I agree totally' was given the index value 5 and the response 'I disagree totally' was given the index value 1. The total index for each individual could vary between 6 and 30. The three independent variables used were gender, area of specialization (Engineering/Business), attitudes to new technology (measured in five levels: innovator, early adopter, early majority, late majority and laggards). The factor analysis performed showed a close inter-relationship (factor score more than 0.5) between the questions. From the multiple regression analysis attempted using the variable 'Attitude' and the three independent variables gender, area of specialization (Engineering/ Business), attitudes to new technology (measured in five levels: innovator, early adopter, early majority, late majority and laggards), there was seen a significant statistical relationship between the attitude index and attitudes to new technology (p-value, $p<0.05$; coefficient of determination, $R^2=3\%$) at 5% significance level. But there was seen no significant relationship ($p>0.1$) between attitude and gender or attitude and area of specialization at even 10% significance level. Figure 2 reveals the distribution of the attitude index in each of the five categories showing the attitude to technology. Clearly the innovators show a closer distribution of the attitude scores between roughly 17 and 27 with the highest

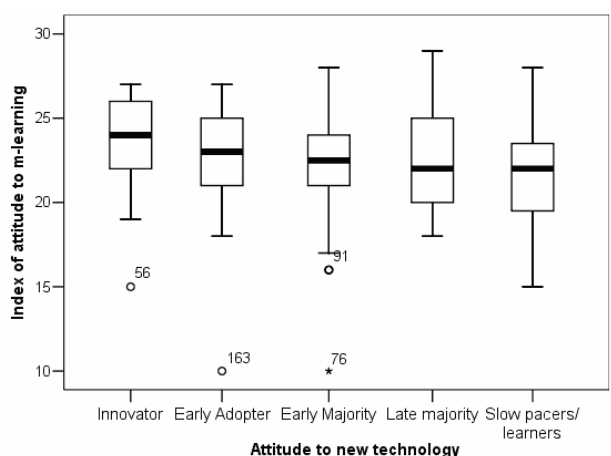

Fig 2. Box plot showing the Attitude Index to m-learning on the Attitude to new technology categories.

minimum and an outlier which is not so far compared to the other four categories.

Under the third objective, through two closed ended questions and one open ended question information was gathered on the most important advantages and disadvantages, from the students' experience with the m-learning system in place in the university. As shown in table 1, the main advantages highlighted were – there was easy access to learning materials/resources, and there was no need to wait for lab or library PCs to be free. The disadvantages highlighted included that the laptop needed to be carried by the student to the school from home. Also the non availability of a laptop caused some to have no access to the wireless network at all. They also expressed concern about bandwidth and speed when many users were connected to the access points simultaneously. The students did not seem to be much aware of the security issues with wireless networks. The student responses were also invited through another closed question regarding how they would like to see mobile learning in the future. Majority of students voiced that they wanted laptops, PDAs (Personal Digital Assistants) and hand phones to be used together for communication and learning. The open ended question was aimed at getting at least two points on mobile learning from the student's point of view in the campus. The qualitative information was content-analyzed and classified into main categories. Students responded that wireless networks were a must for university campuses as it's not possible to offer PCs to all the students. The system was praised as a tool which made the university life less time consuming since students could download necessary information for assignments during anytime they are free. Some voiced that carrying a mobile phone is handy compared to laptops; hence it's convenient to use high end mobile phones for accessing wireless networks. A minority commented that not many could afford to buy a laptop or high end mobile phones to connect to the wireless networks. In general students felt that the wireless network boosted efficiency and effectiveness on both sides (the student and the lecturer) of the learning endeavour.

TABLE I
THE ADVANTAGES WITH M-LEARNING WITHIN A UNIVERSITY CAMPUS.

| Advantages with m-learning | Percentage in favour |
|---|---|
| Easy access to learning resources | 74% |
| Learning is easier as I can chat with online friends to clarify doubts | 43% |
| No need for lab or library PCs to be free | 79% |
| Less virus attack as we can use our own wireless laptop to connect to network | 30% |
| Communication is a lot easier with teachers and friends | 33% |

IV. CONCLUSION

The analysis of student perceptions on m-learning points to the fact that mobile learning is widely embraced by the student community. The majority of students supported the notion that the wireless networks increase the flexibility of access to resources in learning and that they could work independently of available resources like lab or library PCs. The students also were keen to use all sources of m-learning approaches through laptops, palmtops, mobile phones and PDAs so that access to information would be anytime and anywhere.

ACKNOWLEDGMENT